\documentclass[numberedappendix]{emulateapj}
\usepackage{aas_macros}
\usepackage{amsmath,amssymb}
\usepackage{natbib}    
\usepackage{hyperref} 
\usepackage{graphicx} 
\usepackage{color}

\renewcommand{\b}[1]{\boldsymbol{#1}}
\newcommand{\unit}[1]{\,{\rm #1}}

\newcommand{\mean}[1]{\langle #1\rangle}
\newcommand\au{\mbox{\,\textsc{au}}}
\newcommand{\cm}{\unit{cm}}

\newcommand{\g}{\unit{g}}

\newcommand{\K}{\unit{K}}

\newcommand{\kms}{\unit{km\,s^{-1}}}

\newcommand{\msunperyr}{\unit{M_\odot\,yr^{-1}}}

\newcommand{\erg}{\unit{erg}}

\newcommand{\keV}{\unit{keV}}
\newcommand{\s}{\unit{s}}
\newcommand{\yr}{\unit{yr}}
\newcommand{\am}{\mathrm{Am}}
\newcommand{\cs}{c_{\rm s}}
\renewcommand{\d}{\mathrm{d}}
\newcommand{\gr}{\mathrm{gr}}

\newcommand{\VA}{V_{\textsc{a}}} 

\newcommand{\fuv}{\mathrm{FUV}}
\newcommand{\pos}[1]{{\rm#1}^+}

\begin{document}

\title{Wind-driven Accretion in Transitional
  Protostellar Disks}

\author{Lile Wang$^1$ and Jeremy J. Goodman$^1$}

\footnotetext[1]{Princeton University Observatory,
  Princeton, NJ 08544}

\begin{abstract}
  Transitional protostellar disks have inner cavities
  heavily depleted in dust and gas, yet most show signs of
  ongoing accretion, often at rates comparable to full
  disks.  We show that recent constraints on the gas surface
  density in a few well-studied disk cavities imply that the
  accretion speed is at least transsonic.  We propose that
  this is the natural result of accretion driven by
  magnetized winds.  Typical physical conditions of the gas
  inside such cavities are estimated for plausible X-ray and
  FUV radiation fields.  The gas is molecular and
  predominantly neutral, with a dimensionless ambipolar
  parameter in the right general range for wind solutions of
  the type developed by K\"onigl, Wardle, and others.  That
  is to say, the density of ions and electrons is sufficient
  for moderately good coupling to the magnetic field, but
  not so good that the magnetic flux need be dragged inward
  by the accreting neutrals.
\end{abstract}


\section{Introduction}
\label{sec:intro}

Transitional protostellar disks (hereafter TDs) are
deficient in mid-infrared emission
($\lambda\lesssim 10\micron$), implying a dearth of small
dust grains interior to a few to several tens of $\au$
\citep[and references therein] {Skrutskie+etal1990,
  Espaillat+etal2014}. TDs are often supposed to represent
an evolutionary stage intermediate between classical T Tauri
systems, which have full disks, and weak-lined T Tauri
systems, which have little or none.  In this view, disks
disperse from the inside out through some combination of
photoevaporation, viscous evolution, and planet formation
\citep[and references therein] {Marsh+Mahoney1992,
  Hollenbach+etal1994, Clarke+etal2001, Alexander+etalPPVI}.

TDs tend to accrete less than full disks of comparable mass
by factors $\sim3$-$10$ \citep{Najita+etal2015}. The
abundance of small dust in the disk cavity is suppressed by
much larger factors \citep{vanderMarel+etal2015}.  Some TDs,
however, have accretion rates entirely comparable to those
of full disks despite inner gaps of tens of $\au$
\citep{Manara+etal2014}.  Evidently gas somehow crosses the
gap between the outer disk and the star.  Indeed, gas is
detected within dust cavities of accreting TDs via
$\mathrm{H}_2$ fluorescence
\citep{Ingleby+etal2009,Gorti+etal2011} and rovibrational CO
emission \citep{Pontoppidan+etal2008,Salyk+etal2009}.  In
both cases the emission appears to correlate with $\dot M$.
\citet{Owen2016} suggests that there are two types of TDs:
(i) those that are faint at submillimeter wavelengths, have
cavities $\lesssim 10\au$, and accrete at
$\dot M\lesssim 10^{-9}\msunperyr$; (ii) those that are
bright in the submillimeter, have cavities $\gtrsim 20\au$,
and accrete at $\sim 10^{-8}\msunperyr$.  The first type is
consistent with expectations for photoevaporating disks, he
suggests, whereas the latter is not.  The present paper
pertains mainly to class (i)---TDs that accrete rapidly
despite large cavities.

The persistence of accretion suggests that within TD dust
cavities either the ratio of small dust to gas is reduced,
or the velocity of accretion is increased (implying a lower
surface density for a given $\dot M$).
Since quantitative measures of the gaseous surface density
within TD cavities are sparse, many theoretical explanations
for TDs have focussed on the former possibility, i.e. on
mechanisms for altering the abundance of small grains per
unit gas mass, such as photophoresis
\citep{Krauss+Wurm2005}, radiation pressure
\citep{Chiang+Murray-Clay2007}, pressure-induced dust
filtering \citep{Paardekooper+Mellema2006, Rice+etal2006,
  Zhu+etal2012}, or grain coagulation
\citep{Tanaka+etal2005}.

Alternatively or in combination with modified dust,
mechanisms for increasing the radial speed of accretion
($v_{\rm acc}$) within TD cavities have been proposed.
These include enhanced MRI turbulence
\citep{Chiang+Murray-Clay2007,Suzuki+etal2010} and planetary
torques \citep{Varniere+etal2006,Zhu+etal2011,
  Rosenfeld+etal2014}.  At $\dot M\sim 10^{-8}\msunperyr$,
MRI alone cannot reduce the surface density below a few
$\unit{g\,cm^{-2}}$ at relevant radii, which does not render
the disk optically thin with standard ISM dust and would
appear also to violate direct constraints on the gas column
in some systems (\S\ref{subsec:vr}).  Planetary-torque
models, even with multiple planets, have difficulty opening
wide and clean gaps while still maintaining high accretion
rates.

In this paper, we focus on a well-recognized mechanism that
naturally produces rapid accretion: magnetized disk winds.
Despite much work on disk winds since the seminal papers
of \citet{Blandford+Payne1982} and \citet{Pudritz1985},
applications to transitional disks have rarely been remarked
upon.  Notably however, \citet{Combet+Ferreira2008}
envisage a disk with relatively high surface density driven
by turbulent viscosity at large radius, but with a lower
wind-driven surface density within some transition radius
$r_J$.  Even in that paper, the term ``transitional disk''
appears only once in passing.  The authors' motivation
appears to have been mainly theoretical, having to do with
inward concentration of large-scale poloidal magnetic flux
brought about by competition between radial advection and
turbulent diffusion.  The radial redistribution of magnetic
flux is a difficult problem (\S\ref{subsec:flux}), and we
are agnostic as to whether CF08's transition radius $r_J$
can be predicted from first principles.

Our motivation for considering this type of model arose from
other considerations.
First, MRI-driven turbulence in active layers of
disks with total surface densities comparable to the
minimum-mass solar nebula seems only marginally viable at
radii $\sim\mbox{1-10}\au$: such turbulence requires field
strengths approaching equipartition with the local gas
pressure in order to reproduce observed accretion rates
\citep{Bai+Goodman2009,Bai2011}.  Secondly, recent numerical
simulations of MRI turbulence with ambipolar
diffusion often laminarize and produce spontaneous outflows
\citep{Bai+Stone2013,Gressel+etal2015}.  Thirdly,
recent constraints derived from ALMA data on
the gas content within the cavities of a few robustly
accreting TDs demand accretion speeds at least as high as
the gas sound speed (\S\ref{subsec:vr}).

The outline of the paper is as follows.  \S\ref{sec:background}
summarizes the observational evidence for rapid inflow and the
theoretical reasons for expecting wind-driven accretion to be
transsonic.  We also review the importance of ambipolar diffusion, at
least under laminar conditions, for allowing the gas to accrete
without overly concentrating the magnetic flux.  This leads to a
calculation in \S\ref{sec:environment} of the expected degree of
ionization and ambipolar coupling of the gas to the field in the
cavity regions.  It is found that ambipolar diffusion is plausibly in
the Goldilocks range: neither so rapid as to undercut the magnetic
wind torque on the neutrals, nor so slow as to cause the field to be
accreted with the gas.  We write ``plausibly'' because this is a
complex calculation subject to many uncertainties in the chemical
network, dust effects, and radiation field.
\S\ref{sec:discussion-summary} summarizes our findings and directions
for future research.

\section{Dynamical considerations}
\label{sec:background}

\subsection{Constraints on inflow speed}
\label{subsec:vr}

If the surface density of gas in a TD dust cavity were
known, then the inflow speed could be estimated from the
observed accretion rate on the assumption of steady state.
While gas has been detected in a number of TD cavities,
the amount of gas has not often been reliably
quantified.  \citet{Bruderer2013} has argued that CO should
survive photodissociation even if the gas and dust surface
densities are suppressed by several orders of magnitude.
Rotational transitions
trace the total gas content more reliably than rovibrational ones
because of the lower excitation temperatures and critical
densities of the former.  ALMA has now made it possible to resolve TD
cavities in CO rotational transitions.  In a recent study,
\citet[hereafter vdM16]{vanderMarel+etal2016a} have applied
this method to four TDs, using multiple isotopologues of
carbon monoxide ($^{12}$CO, $^{13}$CO, C$^{18}$O) to correct
for optical depth and photodissociation.  Three of their
four systems have quoted accretion rates
$>10^{-9}\msunperyr$.  Of these, DoAr44 is particularly
interesting because the gas surface density ($\Sigma$)
within $16\au$ is inferred to be no more than
$\sim 10^{-4}$ of an extrapolation from the outer disk.
Specifically, in vdM16's model,
$\Sigma(r)\lesssim 5\times10^{-3}
~r_{16}^{-1}\unit{~g~cm^{-2}}$
for $r_{16}\equiv r/(16\au)<1$.  Yet
$\dot M=9\times 10^{-9}\msunperyr$.  The implied accretion
speed is
\begin{equation}
  \label{eq:radial-speed-DoAr44}
  v_{\rm acc} = \frac{\dot M}{2\pi r\Sigma} \gtrsim 0.8\kms.
\end{equation}
This is comparable to the sound speed if the hydrogen is molecular and
its temperature is in equilibrium with the radiation field of the star
($T_{\rm eq}\approx 100\,r_{16}^{-1/2}\K$):
$c_{\rm s}\approx 0.7r_{16}^{-1/4}\kms$.  Unless $\Sigma$ is much less
than vdM16's upper bound, the accretion speed
\eqref{eq:radial-speed-DoAr44} is much less than free fall speed,
$v_{\rm ff}\approx 12\,r_{16}^{-1/2}\kms$.  On the other hand, it is
much larger than can be explained by ``viscous'' accretion mechanisms:
the implied Shakura-Sunyaev viscosity parameter would have to be
$\alpha = \dot M \Omega / 3\pi \cs^2 \Sigma \gtrsim 8$.
Similar conclusions follow for the other two of
\citet{vanderMarel+etal2016a}'s systems that show appreciable
accretion, viz. HD135344B \& IRS48: the accretion speed is
superviscous ($\alpha>1$) but consistent with being subsonic or
transsonic.

\subsection{Wind-driven inflow}

Magnetized winds from thin disks naturally drive much faster
accretion than ``viscous'' mechanisms that conserve the angular
momentum within the disk, such as magnetorotational turbulence.  This
has been noted previously \citep[e.g.][]{Pelletier+Pudritz1992}, but
because it often goes unremarked in works that focus on the launching
or collimation of winds, the reasons are worth reviewing.

Consider an annular control volume
$(z,r,\phi)\in[-h,+h]\times[r_1,r_2]\times[-\pi,\pi]$ in
cylindrical coordinates aligned with the disk, which has
half thickness $\sim h\ll r_1$.  By integrating the mass
flux $\rho\b{v}$ and angular-momentum flux\footnote{We neglect
  Reynolds compared to Maxwell stresses, so that $\b{v}$ can
  be replaced by its temporal and azimuthal mean.}
$r(\rho v_\phi\b{v}-B_\phi\b{B}/4\pi)$ over the surface of
this volume and assuming an axisymmetric statistical steady
state, one easily sees that
 \begin{multline*}
   \left[\dot M_a(r) j(r) +
     \frac{r}{2}\int\limits_{-h}^h
     B_rB_\phi\,dz\right]_{r=r_1}^{r=r_2}\\
   = \int\limits_{r_1}^{r_2}\left(\frac{d\dot
         M_w}{dr}j(r)-\frac{r^2}{2}\left[\vphantom{\frac{1}{2}}B_\phi
       B_z\right]_{z=-h}^{z=h}\right)dr,
 \end{multline*}
 in which $\dot M_a(r)$ is the total mass flow inward past
 radius $r$, $\dot M_w(r)$ is the total outflow of the wind
 from the disk surfaces ($z=\pm h$) within radius $r$, and
 $j(r)=rv_\phi\approx \sqrt{GM_*r}$ is the mean specific
 angular momentum at $r$.  Noting that
 $d\dot M_a/dr=d\dot M_w/dr$ in steady state and taking
 $r_2\to r_1$ leads to a differential statement of the
 conservation of angular momentum:
\begin{equation}
  \label{eq:Jcons}
  r^{-1}\dot M_a\frac{dj}{dr} = -  h\overline{B_r B_\phi}
    -\tfrac{1}{2}r\left[ (B_zB_\phi)^+ - (B_zB_\phi)^-\right],
\end{equation}
where the overbar indicates a vertical average, and $B_i^\pm$ is
shorthand for $B_i(r,\pm h)$.

Thus the advantage of winds over MRI turbulence in driving
accretion is geometrical: the wind stress $B_zB_\phi$ acts on a
surface area $\sim \pi r^2$, whereas the ``viscous'' $B_rB_\phi$
stress acts on a much smaller area $\sim 2\pi r h$.
If the radial and vertical
components of the magnetic field have comparable strength,
then the former exert a torque larger than the latter by a factor
$\sim r/h \sim \Omega r/\cs \gg 1$.
With the usual symmetry
$(B_z,B_r,B_\phi)^+=(B_z,-B_r,-B_\phi)^-$, the accretion
speed becomes
\begin{equation}
  \label{eq:vfromB}
  v_{\rm acc}\approx - \frac{B_zB_\phi^+}{\pi\Sigma\Omega}\ .
\end{equation}
If all three components of the field are comparable, then
this becomes $v_{\rm acc}\approx \VA^2/\Omega h$, where
$\VA^2=B_z^2/4\pi\rho_0$ is the Alfv\'en speed based on
conditions at the midplane, and $h\equiv\Sigma/2\rho_0$ is
the effective half thickness of the disk.  If this thickness
is determined by a balance between gas pressure and the
tidal field, then $\Omega h\sim\cs$, so that the Mach number
of accretion is $\sim (\VA/\cs)^2$.  Hence winds with
near-equipartition fields should drive transsonic accretion.

In the semi-analytic wind models developed by
\citet{Konigl1989}, \citet{Wardle+Konigl1993}, and
\citet{Li1996}, $v_{\rm acc}$ is in fact typically
comparable to the sound speed, or even a few times
larger.\footnote{The horizontal components of field, which
  vanish at the midplane, compress the thickness $h$ to less
  than its tidal value, strengthening the conclusion of the
  previous paragraph.}  This is also true of numerical
simulations that treat the disk structure explicitly and
approach a steady state
\citep{Casse+Keppens2002,Zanni2007,Tzeferacos+etal2009}.

\subsection{Advection of flux}
\label{subsec:flux}

Despite their greater efficiency at removing angular
momentum from accreting matter, wind models are generally
more complicated to use than turbulent/viscous ones.  The
magnetic geometry of the wind must be solved for globally,
and the loading of the field lines with matter launched from
the disk surface must be consistent with the physical state
of the gas at that surface.  Wind dynamics cannot easily be
reduced to a single dimensionless parameter comparable to
the Shakura-Sunyaev turbulent-viscosity parameter $\alpha$.

In steady state, matter accreting
through the disk must be able to cross field lines so that
magnetic flux is not dragged with it.  Otherwise,
the radial concentration of flux will exert an outward force
on the disk, producing subkeplerian rotation and possibly
choking off the accretion or at least slowing it down
\citep{Bisnovatyi-Kogan+Ruzmaikin1976,Narayan+etal2003}.
Indeed, the tendency to subkeplerian rotation might prevent
the formation of protostellar disks in the first place
\citep[and references therein]{Li+etal2014}.
The wind models of K\"onigl and collaborators invoke
ambipolar diffusion to separate the accreting neutral gas
from the magnetic flux. This process is characterized by
the ratio of the neutral-ion collision time to the orbital
time,
\begin{align}
  \label{eq:ambipolar}
  \am &\equiv \dfrac{\tau_\mathrm{ni}^{-1}}{\Omega}\nonumber\\
      &\sim 0.56\ \left( \dfrac{n_e}{\cm^{-3}} \right)
        \left( \dfrac{r}{16~\au} \right)^{3/2}
        \left( \dfrac{M_*}{M_\odot} \right)^{-1/2}\ .
\end{align}
Here we have evaluated the parameter $\am$ for the surface-density profile
fitted by vdM16 to DoAr44.  It will be shown in
\S\ref{sec:environment} that the electron density at the
midplane within the disk cavity is plausibly such that
$\am\sim 1\mbox{-}10$.

The ambipolar drift velocity between the neutral and charged
species when the latter are tied to the field is
\begin{equation}
  \label{eq:vdrift}
  \b{v}_{\rm drift}\approx \mathrm{Am}^{-1}
  \frac{\b{B\times}(\b{\nabla\times B})}{4\pi\rho\Omega}\,,
\end{equation}
and $v_{\rm drift}\approx v_{\rm acc}$ if the field is stationary.
The reduction in orbital
velocity due to radial magnetic force is
\begin{equation}
  \label{eq:subkep}
  \Delta v_\phi \approx
  \b{\hat e}_r\cdot\frac{\b{B\times}(\b{\nabla\times B})}
  {8\pi\rho\Omega}\,.
\end{equation}
Hence $v_{{\rm drift},\,r}/\Delta v_\phi\approx 2\mathrm{Am}^{-1}$.
In order that $v_{{\rm drift},\,r}\sim \cs$ while
$\Delta v_\phi\ll \Omega r$, it is therefore necessary that
$\mathrm{Am}\ll 2\Omega r/\cs\sim 20$.  On the other hand,
$\am\gtrsim 1$ is needed for adequate coupling.  We are not aware of a
feedback mechanism to ensure that this rather narrow parameter
range is realized at the midplane in the cavity.

Following the methods presented in \S\ref{sec:model-setup},
``fiducial'' values of the radiation field and other physical
parameters do indeed suggest that $\am$ is in the desired range or
close to it.  However, this is likely not to be the last word on the
subject because of the complex microphysics involved in
the calculation and its sensitivity to input parameters such as the
X-ray flux and dust abundance.  {\it A priori}, if $\am$ lies outside the desired
range, it seems more likely to be too large than too small on the grounds
that (i) surely $\am\gtrsim 1$ in the active layers of full disks and
the outer parts of transitional disks, else no magnetic mechanism,
whether relying on MRI or winds, could be effective in driving
accretion there; and (ii) the lower surface density of the cavity gas
and greater proximity to the sources of ionization might be expected
to raise $\am$. 

The problem of accretion of magnetic flux in ideal MHD is
not new.  When accretion is assumed to be driven by
turbulent ``viscosity'', the concern is often opposite to
that expressed here: namely, that turbulent diffusivity
might allow the poloidal flux to escape too easily
\citep{vanBallegooijen1989, Lubow+etal1994,
  Spruit+Uzdensky2005}.  In steady-state wind models based
on near-ideal MHD, just enough turbulent
diffusivity is invoked to allow the flux to be stationary as
the gas accretes \citep[and references
therein]{Blandford+Payne1982, Ferreira+Pelletier1995,
  Zanni+etal2007}.  The questions then become (i) What is
the origin of the turbulence?  (ii) How is the turbulent
diffusivity regulated so that the magnetic flux neither
accretes nor escapes?

A possible answer to question (i) above is MRI, whose
turbulent diffusivity has been measured by several authors
\citep{Lesur+Longaretti2009, Guan+Gammie2009,
  Fromang+Stone2009}.  But MRI tends to be inhibited or
suppressed by poloidal fields as strong as are found in many
wind models \citep{Salmeron+etal2007, Bai+Stone2013,
  Gressel+etal2015}.
If the field \emph{were} to become too strong to allow MRI,
it is unclear what instability would arise to cause that
field to reconnect or diffuse so as to allow MRI to return:
that is, we lack an answer to question (ii) above.
Nonaxisymmetric interchange instabilities are possible but
should occur only if $d(B_z/\Sigma)/dr<0$,
\citep{Spruit+etal1995, Stehle+Spruit2001,
  Igumenshchev2008}, and while the interchange instability
can rearrange the radial profile of $B_z$, it does not by
itself alter $B_z/\Sigma$ in a lagrangian sense.  It may not
be effective in reconnecting the horizontal components of
magnetic field.  More promising in this regard is the
recently much-discussed plasmoid instability of current
sheets \citep{Loureiro+etal2007,Loureiro+Uzdensky2016}, but
this instability has not yet been studied in the presence of
a field component perpendicular to the sheet ($B_z$ here).

\section{Physical state of the cavity gas}
\label{sec:environment}

As the discussion above makes clear, it is crucial to
understand the degree of ionization of the gas in TD
cavities.  In this and in the following section, we assume a
low surface density compatible with transsonic accretion,
and then calculate the resulting ionization fraction, ohmic
diffusivity, and ambipolar parameter.

\subsection{First estimate}
\label{sec:chem-estimates}

A simple analytic estimate of the electron density can be obtained by
balancing X-ray ionization and radiative recombination of hydrogen.
The surface density of the gas in the cavity is so low that we take it
to be optically thin to the X-rays.

Recent observations of TDs indicate a typical X-ray
luminosity $L_X\sim 10^{30}\unit{erg\,s^{-1}}$, and a
characteristic photon energy $\sim 1~\keV$
\citep{Kim+etal2013}. Balancing photoinization of molecular
hydrogen against radiative recombination to atomic hydrogen
\citep[see][]{Draine_book}, one obtains
\begin{align}
  \label{eq:ionization-sol}
    n_e & \sim 40~\cm^{-3}\
    \left( \dfrac{x_e}{3\times 10^{-6}} \right)
    \left( \dfrac{r}{16~\au} \right)^{-2}
    \left( \dfrac{\mean{h}}{r} \right)^{-1}\,,\nonumber\\
    x_e & \sim 3\times 10^{-6}
    \left( \dfrac{T}{10^2~\K} \right)^{0.46}
    \left( \dfrac{\mean{h\nu}}{\keV} \right)^{-2}\,.
\end{align}

This rough estimate would suggest that ambipolar diffusion is
inefficient, i.e. $\am\gg1 $ in eq.~\eqref{eq:ambipolar}.  But the
estimate neglects important recombination processes involving
molecules and dust grains.


\subsection{Improved estimate: Method}
\label{sec:model-setup}

For a better estimate, we take into account multiple
chemical species in addition to hydrogen, including a
simplified treatment of dust grains, and far-UV (FUV) as well as
X-ray ionization.

The calculations are done in spherical polar coordinates, dividing the
gas into 100 radial zones spaced logarithmically in {\it cylindrical}
radius from $r=1\au$ to $r=50\au$, and 20 latitudinal zones spaced
linearly from $\theta=0$ (midplane) to $\theta = 0.2$.  Radiative
transfer of the ionizing photons is calculated by integrating along
radial rays, so that the calculations are effectively one-dimensional,
except for some implicit scattering of the X-rays after they strike
the disk surface.  The surface mass density within the cavity is taken
as $\Sigma \sim 5\times 10^{-3}~r_{16}^{-1}~\g~\cm^{-3}$ following
\S\ref{subsec:vr}.  To allow for a finite angle between the ionizing
rays and the disk surface [e.g. \citet{Perez-Becker+Chiang2011c}], we
consider a flared disk with $h\propto r^{5/4}$ and $h=0.1 r$ at
$r=16\au$ ($r_{16}=1$).  Hydrogen nuclei are distributed as
\begin{equation*}
  n_{\textsc{h}}(r,z) = n_0r_{16}^{-2.25}
  \exp[-(r \tan\theta)^2/(2h^2)] \ ,
\end{equation*}
with $n_0=10^{8\pm1}\unit{cm^{-3}}$. The gas is assumed
vertically isothermal with radial temperature profile
$T=T_0r_{16}^{-1/2}$ and $T_0 = 100~\K$, $30~\K$ or
$300~\K$.

Ionizing photons emanate from the origin $r=0$ with total
luminosities $L_X = 10^{30\pm1}~\erg~\s^{-1}$ and
$L_\fuv = 10^{30\pm1}~\erg~\s^{-1}$.  FUV photons are taken
to be monochromatic with $h\nu=12\unit{eV}$ Extreme UV
photons ($h\nu>13.6\unit{eV}$) are ignored on the grounds
that their penetration depths are negligible.
Even FUV photons have rather little effect on the ionization
balance at depth in the disk because of our restriction to
purely radial propagation:  The radial column density that
must be crossed to reach the midplane at $r\gtrsim 1\au$ is
$N_{\mathrm{H}_2} \sim 10^{24}~\cm^{-2}$, falling to
$\sim 10^{22}~\cm^{-2}$ at 1-2 scale heights ($h$) off the
plane.

X-ray penetration is calculated via the fit by
\citet{Bai+Goodman2009} to the Monte-Carlo results of
\citet{Igea+Glassgold1999}, and hence implicitly includes
some scattering and secondary effects.  Following
  \citet{Bai+Goodman2009}, all ionizations are attributed to
hydrogen and helium for the purpose of solving the chemical network;
however, the total X-ray cross section per gram of gas takes into
account a solar abundance of oxygen and other metals.
X-ray photons are taken to be monochromatic with
$h\nu = 3~\keV$.

Time-dependent rate equations
\begin{equation}
  \label{eq:chem-ode}
  \dfrac{\d x^i}{\d t} = \sum_{j,k} A^i_{\;jk}x^j x^k +
  \sum_{j}B^i_{\;j}x^j\ ,
\end{equation}
are integrated to obtain the concentrations $\{x^i\}$ of the
various species. The rate coefficients $\{A^i_{\;jk}\}$ of
two-body reactions are taken from the \verb|UMIST|
astrochemistry database \citep{UMIST2013}, and the
$\{B^i_{\;j}\}$ of photoionization and photodissociation
reactions are caluculated with the schemes elaborated
earlier in this section. Nine abundant elements (H, He, O,
C, N, S, Si, Mg, and Fe) are included at IN06 abundances
together with compounds totalling 175 species
\citep[following Table A.1 of][with $\pos{H}$
added]{Ilgner+Nelson2006}.

Since grains can be important for electron mobility and
recombination, the following reactions are included (X
stands for atomic or molecular species)
\begin{itemize}
\item $\pos{X} + \gr \rightarrow \mathrm{X} + \gr^+ $;
\item $\pos{X} + \gr^- \rightarrow \mathrm{X} + \gr $;
\item $e^- + \gr \rightarrow \gr^- $;
\item $e^- + \gr^+ \rightarrow \gr $.
\end{itemize}
Higher grain charges are ignored because we assume small
grains (see below).  The rates of these reactions are
estimated following \citet{Ilgner+Nelson2006} and
\citet{Bai+Goodman2009}, with reference to
\citet{Draine+Sutin1987} for collision rates and to
\citet{Nishi+etal1991} for electron sticking probabilities.
We set the dust-to-gas mass ratio to $\sim 10^{-4}$,
the density within grains to $3\unit{g\,cm^{-3}}$, and the
grain radius to $10^{-3}~\micron$.  These choices bound the
plausible area of grain surface from above, and hence lead
to a conservative lower bound on $n_e$.  Grain-assisted
molecular hydrogen formation is estimated following the
prescription of \citet{Bai+Goodman2009}.

Equations \eqref{eq:chem-ode} are integrated to steady state
cell by cell, ignoring advection of species between cells
(except ionizing photons), because the local equilibration
timescales are usually short compared to advection times, as
discussed below (\S\ref{subsec:results}).

\subsection{Second estimate: Results}
\label{subsec:results}

\begin{figure}
  \centering
  \includegraphics[width=3.1in, keepaspectratio]
  {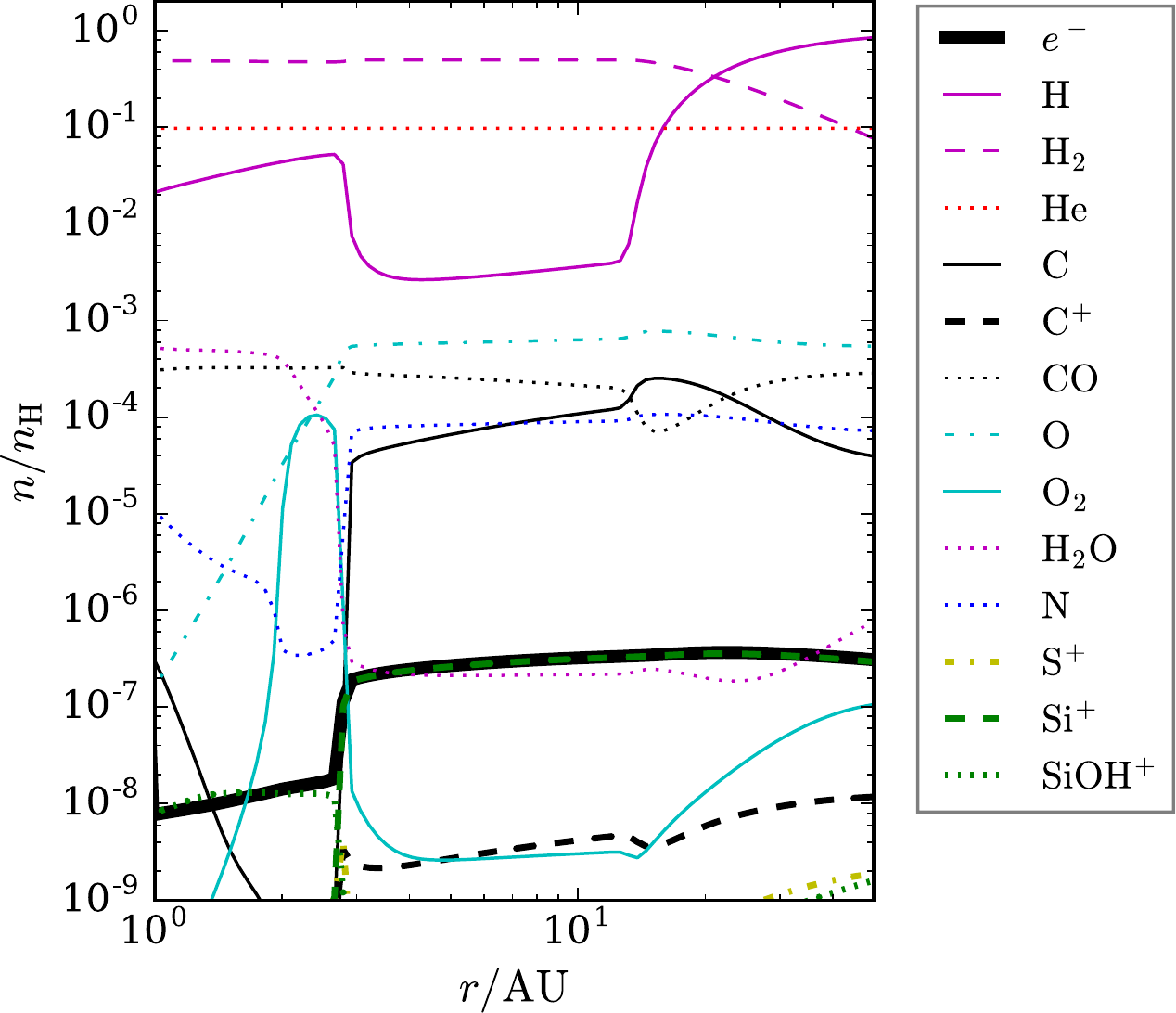}
  \caption{Relative abundance of significant neutral
    ($n/n_\mathrm{H} > 10^{-5}$) and ionic
    ($n/n_\mathrm{H} > 10^{-8}$) species.  The relative
    abundance of electrons is shown by the heavy magenta
    curve; other species as noted in the legend.}
  \label{fig:chem_fiducial}
\end{figure}

Figure~\ref{fig:chem_fiducial} plots the abundance at the midplane
($\theta = 0$) of major species relative to hydrogen for the fiducial
model, in which $L_X = L_\fuv = 10^{30}~\erg~\s^{-1}$,
$n_0 = 10^8~\cm^{-3}$, $T_0=100~\K$,
and $r_\mathrm{dust}= 10^{-3}~\micron$.  The electron abundance
depends depend rather weakly on radius, while the dominant positive
ions are $\mathrm{Si}^+$ and $\mathrm{SiOH}^+$. Photoionization of
helium and oxygen are competitive with that of molecular hydrogen as a
source of free electrons.

For the fiducial case, the timescale on which $n_e$ equilibrates is at
least one order of magnitude shorter than the local accretion
timescale $\tau_\mathrm{acc} \equiv r / v_\mathrm{acc}$ based on
eq.~\eqref{eq:radial-speed-DoAr44}, justifying our neglect of
advection. This is not always true, however, especially in cases
without dust.


\begin{figure*}
  \centering
  \includegraphics[width=7.1in, keepaspectratio]
  {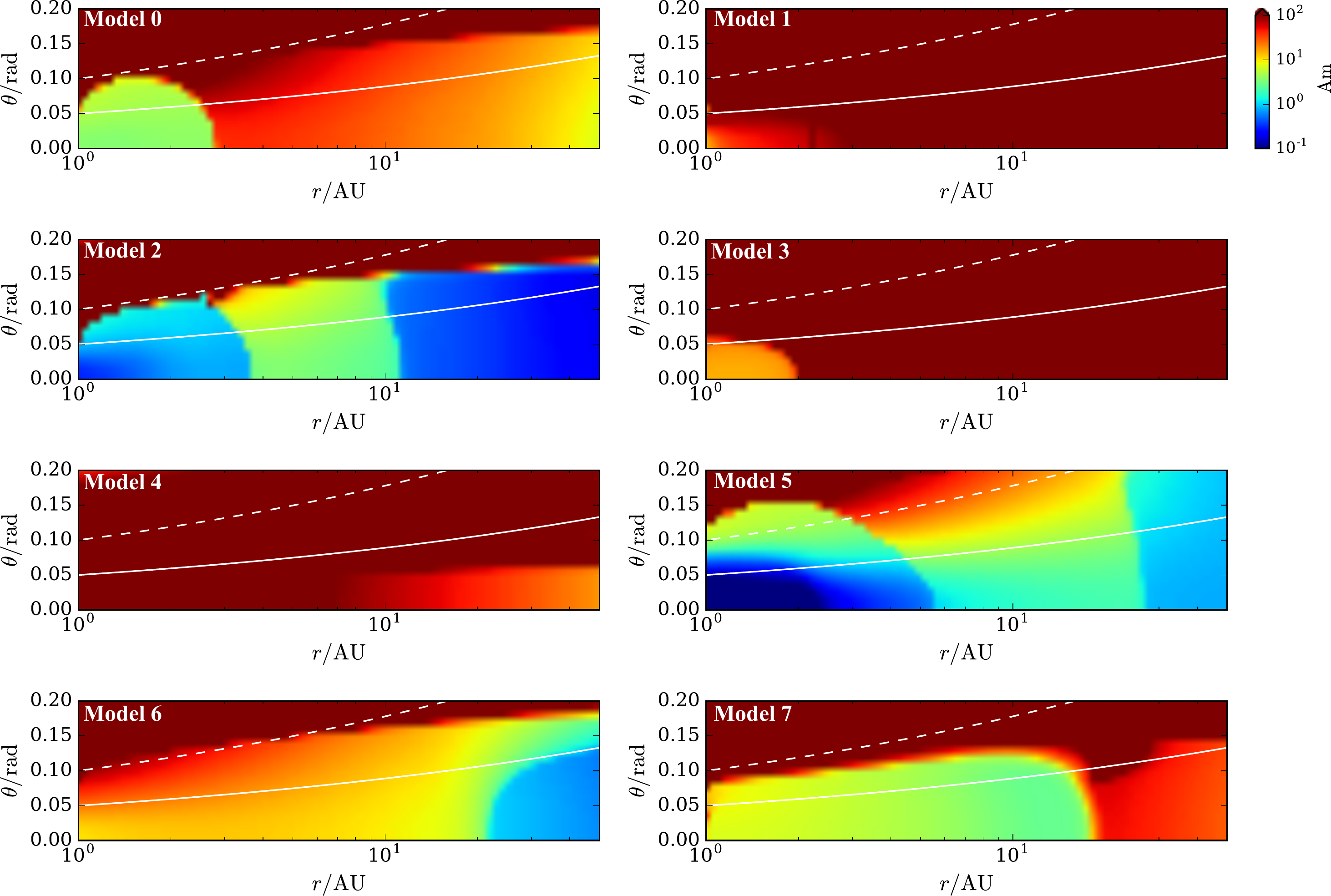}
  \caption{Ambipolar parameter $\am$ vs.
    cylindrical radius $r$ and latitude 
    $\theta$.   Each panel corresponds to one of the
     models in Table~\ref{tab:model-par}, as marked.
    White curves are loci of constant
    gaussian scale height: $h$ (solid), and $2h$ (dashed).}
  \label{fig:Am_2D}
\end{figure*}

\begin{table}
  \centering
  \caption{Model number and parameters} 
  \vspace{0.5cm}
  \begin{tabular}{cc}
    \hline
    Model No. & Description \\
    \hline
    0 & Fiducial \\
    1 & No dust \\
    2 & $L_X = 10^{29}~\erg~\s^{-1}$ \\
    3 & $L_X = 10^{31}~\erg~\s^{-1}$ \\
    4 & $n_0 = 10^{7}~\cm^{-3}$ \\
    5 & $n_0 = 10^{9}~\cm^{-3}$ \\
    6 & $T_0 = 30~\K$ \\
    7 & $T_0 = 300~\K$ \\
    \hline
  \end{tabular}
  \label{tab:model-par}
\end{table}

We calculate seven additional models, each differing from
the fiducial one in one parameter
(Table~\ref{tab:model-par}).  Following
eq.~\eqref{eq:ambipolar}, Fig.~\eqref{fig:Am_2D} exhibits
the profiles of $\am$ in the $rz$ plane for all eight
models.

The electron concentration and therefore $\am$ increase as
X-ray luminosity increases, and as density decreases: the
latter is contrary to the trend for radiative hydrogen
recombination and is due to molecular ions.  The dependence
on temperature is relatively complicated and subtle, as
temperature affects the rates of radiative, dissociative and
dust-assisted recombination in different ways. Comparison
between Models 0 and 1 illustrates the importance of small
dust.  Eliminating dust-assisted recombination raises the
electron density by one to two orders of magnitude.

The fiducial model has a relatively low-$\am$ region that
extends radially to $\sim 3~\au$, and vertically to 1-2
scale heights ($h$).  Similar regions exist in Models 2, 5
and 7, and less clearly in Models 1 and 3. The exponential
attenuation of X-rays with increasing column density
actually lowers the ionization parameter
$\xi \equiv \zeta / n_\mathrm{H}$ at smaller radii. The
boundary of the low-$\am$ regions corresponds to a
transition in the dominant ion from $\mathrm{Si}^+$ at
relatively high $n_e$, to $\mathrm{SiOH}^+$ (which has a
larger dissociative-recombination cross section) at lower
$n_e$.

We have used these results for the physical state of the gas
to recompute the local wind model of
\citet{Wardle+Konigl1993}.  As already noted, their model
requires $\am\sim O(1)$ near the midplane, so that the
coupling of the neutrals to the magnetic field is efficient
enough to drive a wind but not so perfect as to accrete the
magnetic flux.  Fig.~\eqref{fig:Am_2D} shows that
$0.3\lesssim\am\lesssim30$ at $r\lesssim 10\au$ in most of
our models, excepting those with rather high ionization
parameter or without dust.  Quantitatively, we find that the
radial velocity of the neutral gas at the midplane varies
from $\sim 10^{-1}$ to $\sim 10^0c_s$, and the accretion
rate from $10^{-9}$ to $10^{-8}~M_\odot/\yr$.  These latter
are generally consistent with observed accretion rates in
transitional disks.
  
\section{Discussion and Summary}
\label{sec:discussion-summary}

We have seen that magnetized winds can rather naturally
explain the combination of low gas and dust surface density
and relatively robust accretion in transition-disk cavities.
From a theoretical point of view, the suggestion is all the
more natural because of previous work demonstrating the
likelihood of \emph{photoevaporative} winds on the one hand
\citep{Hollenbach+etal1994, Clarke+etal2001,
  Alexander+Clarke+Pringle2006, Owen+etal2010}, and the need
for net poloidal magnetic flux to sustain magnetorotational
turbulence in the minimally ionized parts of protostellar
disks, on the other \citep{Fleming+etal2000,
  Oishi+MacLow2011, Flock+etal2012, Bai+Stone2011,
  Simon+etal2013a, Simon+etal2013b}.  A thermally-driven
outflow threaded by magnetic lines is bound to exert a
torque on the gas remaining in the disk and drive some
accretion, though just how much will require detailed
modeling of a sort not attempted in the present work.

If the whole vertical column of the disk were to accrete at
$v_{\rm acc}\sim\cs$, then the surface density corresponding
to observed accretion rates would be much less than what is
inferred observationally, e.g. from submillimeter
observations \citep{Beckwith+Sargent1993,Andrews+etal2013}.
Thus in the parts of TDs exterior to their cavities, as well
as in full (non-transitional) T~Tauri disks, either
accretion is driven turbulently (and hence relatively
slowly), or else it is again driven by a magnetized wind,
but one that couples to only a small fraction of the
vertical column where FUV and X-ray photons sufficiently
ionize and heat the gas \citep{Bai2016}.

Not addressed here is the \emph{cause} of the transition
between slow/layered accretion at large radii and fast,
wind-driven, full-column accretion within the cavity.
Perhaps, as proposed by \citet{Combet+Ferreira2008}, this
has to do with a critical value of magnetization brought
about by advection of poloidal flux. But as discussed in
\S\ref{subsec:flux}, the advection and diffusion of magnetic
flux in a turbulent disk is an unsolved problem.  Also, if
this is the explanation, why aren't all T~Tauri disks
transitional?  Perhaps indeed giant planets are implicated,
but rather than producing the entire cavity by gravitational
torques alone, such planets trigger a local change in the
mass-to-flux ratio by stemming the inflow of gas near the
midplane (which feels little magnetic torque) but not the
inflow of the more ionized surface layers (which are more
strongly magnetically driven, whether by winds or MRI).  In
any case, one should entertain the possibility that
transitional disks---or at least those that have accretion
rates comparable to those of full disks---may not be an
evolutionary phase that all disks pass through immediately
before complete dispersal.

\vspace*{10pt}

This work was supported by  NASA Origins of Solar Systems grant
NNX10AH37G.

%
%
\bibliography{transit}
\bibliographystyle{apj}
%
%
\end{document}